\documentclass[showpacs,aps,pre,superscriptaddress,twocolumn]{revtex4}

\usepackage{graphicx}
\usepackage{amsfonts,dsfont}
\usepackage{color, verbatim}
\usepackage{array}
\usepackage{algorithm, algpseudocode,boldline}

\algdef{SE}[DOWHILE]{Do}{doWhile}{\algorithmicdo}[1]{\algorithmicwhile\ #1}%

\begin{document}
\title{2D solutions of the hyperbolic discrete nonlinear Schr\"odinger equation}
\author{J. D'Ambroise}
\affiliation{ Department of Mathematics, Computer \& Information Science, State University of New York (SUNY) College at Old Westbury, Westbury, NY, 11568, USA; dambroisej@oldwestbury.edu}
\author{P.G. Kevrekidis}
\affiliation{Department of Mathematics and Statistics, University of Massachusetts,
Amherst, MA, 01003, USA; kevrekid@math.umass.edu}

\begin{abstract}
  We derive stationary solutions to the two-dimensional
  hyperbolic discrete nonlinear  Schr\"odinger (HDNLS) equation by starting from the anti-continuum limit and extending solutions to include nearest-neighbor interactions in the coupling parameter.
  We use pseudo-arclength continuation to capture the relevant branches
  of solutions and explore their corresponding stability and dynamical
  properties (i.e., their fate when unstable).
  We focus on nine primary types of solutions: single site, double site in- and out-of-phase, squares with four sites in-phase and out-of phase in each of the vertical and horizontal directions, four sites out-of-phase arranged in a line horizontally, and two additional solutions having respectively six and eight nonzero sites.  The chosen configurations are found to merge into four distinct
  bifurcation events.
  We unveil the nature of the bifurcation phenomena and identify
  the critical points
  associated with these states and also explore the consequences of the termination of the branches on the dynamical phenomenology of the model.
\end{abstract}

\maketitle

\section{Introduction}

The hyperbolic nonlinear Schr{\"o}dinger equation is a model
of increasing interest both in applied mathematics and in
theoretical/experimental physics~\cite{SuSu,CoDiTri,CoTri,GHS1,GHS2,GHS3}
as it arises in a diverse host of
physical applications. Among others, one can mention
as specific examples deep water waves~\cite{abseg,zakh} and cyclotron
waves in plasmas~\cite{sen,myra}, although the equation has been
also quite popular in nonlinear optics. Within the latter,
the examination of normally dispersive (quasi-discrete)
optical waveguide arrays~\cite{Drou,Lah} has offered a framework
for the study of optical pulses. Additionally, the nonlinear,
experimentally accessible X-wave structures~\cite{trillo1,trillo2}
(but also more elaborate states including dark-bright~\cite{hay}
or vortex-bright~\cite{efr} solitary waves) have motivated
its theoretical and numerical study. More recent efforts have also
seen the development of methods based on hyperbolic coordinates
to study the standing waves of the HNLS~\cite{zeng},
a consideration of its universal asymptotic regime for
a wide range of initial conditions~\cite{rumanov}, as well as
the analysis of its profile decomposition in different mass-critical
and supercritical cases~\cite{dodson}; see also references therein.

While these extensive studies have addressed numerous aspects
of the continuum HNLS model, we are not aware of any efforts
considering the (genuinely) discrete aspects of the model, the
so-called hyperbolic discrete nonlinear Schr{\"o}dinger or
HDNLS equation.
This is an interesting endeavor on a number of counts.
On the one hand, the elliptic variant of the discrete problem is
quite well understood (see, e.g., the monograph~\cite{dnls}),
hence, it is conceivable that some of the corresponding analytical
and numerical techniques
may be adapted to the present setting. In fact, there exists
a so-called staggering transformation $u_{n,m}=(-1)^n w_{n,m}$
(where $w_{n,m}$ is the solution of the elliptic problem) which
can convert the former to the latter. However,  we will not use
this approach here, given that this transformation becomes
singular in the continuum limit. Instead, we note that the
phenomenology of the HDNLS model is of interest not only
given its consideration as a numerical scheme for the continuum
HNLS, but also because some of the applications may bear a(n
at least partially) discrete character, as discussed, e.g.,
in~\cite{Drou,Lah}.

There are additional characteristics that
add to the appeal of the  HDNLS model.
For instance, in the so-called anti-continuum (AC)
limit of vanishing coupling between adjacent nodes, {\it any}
stationary configuration is ``permissible'' as is known also
for the elliptic case~\cite{dnls}. However, in the continuum
limit, on the other hand, the work of~\cite{GHS3} established
that there are no nontrivial standing wave solutions.
This implies that {\it all} the solutions initiated
at the AC limit must terminate at some point prior to
reaching the continuum i.e., at some finite value of the
coupling strength. This is fundamentally different from the
standard elliptic DNLS case, where solitary waves, and even
vortical solutions may persist in the continuum limit~\cite{SuSu}.
Thus, it is of interest to explore the bifurcations through which
these branches terminate and to classify the dynamical behavior
of the model prior to, as well as past the corresponding critical points.
It is the aim of the present work to address a number of these issues
for some of the most fundamental (one- and few-site) configurations
of the HDNLS model.

Our presentation will be structured as follows: In section II, we will
explore theoretical aspects of the existence (via solvability
conditions) and stability theory (linearizing around the equilibrium
configurations and exploring the corresponding spectrum). Then,
numerical computations will be used in section III to corroborate
the analytical existence/stability results and direct numerical
simulations will help us determine the fate of such waveforms
when unstable (or when they may not exist closer to the continuum limit).
Finally, in section IV,
we will summarize our findings and present our conclusions, as well
as a number of challenges towards future work.

\section{Model}

We consider the HDNLS equation for $u_{n,m}(z)$ as follows
\begin{equation}
i \frac{du_{n,m}}{dz}+\epsilon \Delta_H u_{n,m} +|u_{n,m}|^2 u_{n,m}=0 \label{dyneq}
\end{equation}
where $\Delta_H u_{n,m} = u_{n,m+1}+u_{n,m-1}-u_{n+1,m}-u_{n-1,m}$
stands for the hyperbolic operator, i.e., a discretization
of $u_{xx}-u_{yy}$ with unit spacing, while $\epsilon$ is the nearest neighbor coupling parameter. In the context of this being a(n isotropic)
discrete approximation
to the continuum problem, one should think of $\epsilon=1/\Delta x^2$,
where $\Delta x$ is the spacing between adjacent lattice nodes in both
directions.
The indexing $n$ represents the discrete vertical direction and $m$
the horizontal one.
Setting $u_{n,m}(z)=\phi^{(\epsilon)}_{n,m}e^{i\mu z}$ we obtain the stationary equation $F^{(\epsilon)}_{n,m}=0$ for
\begin{equation}
F^{(\epsilon)}_{n,m}(\phi) \stackrel{def.}{=} (\mu-|\phi^{(\epsilon)}_{n,m}|^2)\phi^{(\epsilon)}_{n,m}-\epsilon  \Delta_H \phi^{(\epsilon)}_{n,m}.
\label{stateq}
\end{equation}
We can then seek standing wave solutions with frequency $\mu$, by
solving the algebraic set of Eqs.~(\ref{stateq}).

\subsection{Existence of Solutions}
In the $\epsilon = 0$ anti-continuum limit, the values of $\phi^{(0)}_{n,m}$ for each site $\{n,m\}$ can be chosen independently from each other since the nearest-neighbor coupling parameter is zero.  Localized solutions are thus found by specifying $\phi^{(0)}_{n,m}=0$ for most sites $\{n,m\}$.  For a few nonzero sites we set  $\phi^{(0)}_{n,m} = e^{i\theta_{n,m}}$ with $\mu = 1$ and $\theta_{n,m}\in\{ 0, \pi\}$.  Table I lists some possible solutions for $\phi^{(0)}_{n,m}$ and a naming convention for each example configuration. 

We will use the following general notation.  Let $\mathcal{S}=\{ \phi^{(0)}_{n_1,m_1}, \phi^{(0)}_{n_2,m_2}, ..., \phi^{(0)}_{n_d,m_d}\}$ for $d\in\mathds{N}$ represent an enumeration of the nonzero sites of the initial $\epsilon = 0$ configuration, and let $\vec{\theta}={\rm arg}(\mathcal{S})\in [0,2\pi]^d$ represent a vector whose elements are the arguments of elements of $\mathcal{S}$.  For simplicity we enumerate the nonzero sites in a natural way with the top-most left nonzero site corresponding to the first index.
Notice that the configurations listed in Table I are not necessarily closed loops, but when they are we enumerate from the top left then counterclockwise.    It will be convenient to denote $\vec{\delta}_L$ as a vector whose components are either $0$ or $1$ corresponding to whether the left neighbor (when considered on the full two-dimensional grid) $\phi^{(0)}_{n_j,m_j-1}$ of each element of $\mathcal{S}$ is zero or nonzero.  Similarly define $\vec{\delta}_R, \vec{\delta}_T, \vec{\delta}_B$ corresponding to whether the right ($\phi^{(0)}_{n,m+1}$), top ($\phi^{(0)}_{n-1,m}$), and bottom ($\phi^{(0)}_{n+1,m}$) neighbors of each element of $\mathcal{S}$ are zero or nonzero on the 2D grid.  Finally, let $\vec{\theta}_L, \vec{\theta}_R, \vec{\theta}_T, \vec{\theta}_B\in [0,2\pi]^d$ denote the arguments of the corresponding nonzero nearest neighbors to each element of $\mathcal{S}$ (with the subscripts having the same neighbor designation as above).  Note that since $\mathcal{S}$ contains all of the nonzero elements of $\phi^{(0)}_{n,m}$, the vectors $\vec{\theta}_\star$ are permutations of $\vec{\theta}$.

  \begin{table}
      \begin{tabular}{cc}
      \multicolumn{2}{c}{ \underline{Branch 1}}\\
\underline{Name} &  \underline{Sites}\\
\\
1s &   \begin{tabular}{|c|}\hline + \\  \hline  \end{tabular}  \\
\\
2i-horz &   \begin{tabular}{|>{\centering}p{0.25cm}|>{\centering}p{0.25cm}|}\hline + &  + \tabularnewline \hline  \end{tabular}   \\
\\
 4o-vert &  \begin{tabular}{|>{\centering}p{0.25cm}|>{\centering}p{0.25cm}|}\hline + & + \tabularnewline \hline  -- & -- \tabularnewline \hline  \end{tabular}
      \end{tabular}
      \hspace{.25in}
      \begin{tabular}{cc}
      \multicolumn{2}{c}{ \underline{Branch 2}}\\
\underline{Name} &  \underline{Sites}\\
\\
2o-horz &   \begin{tabular}{|>{\centering}p{0.25cm}|>{\centering}p{0.25cm}|}\hline + &  -- \tabularnewline \hline  \end{tabular}   \\
\\
4o-line &   \begin{tabular}{|>{\centering}p{0.25cm}|>{\centering}p{0.25cm}|>{\centering}p{0.25cm}|>{\centering}p{0.25cm}|}\hline + & + &  -- & -- \tabularnewline \hline  \end{tabular}   \\
\\
\\
\\
      \end{tabular}

\vspace{.1in}

        \begin{tabular}{cc}
      \multicolumn{2}{c}{ \underline{Branch 3}}\\
\underline{Name} &  \underline{Sites}\\
\\
4i-sqr &    \begin{tabular}{|>{\centering}p{0.25cm}|>{\centering}p{0.25cm}|}\hline + & + \tabularnewline \hline  + & + \tabularnewline \hline  \end{tabular}   \\
\\
8s &  \begin{tabular}{|>{\centering}p{0.25cm}|>{\centering}p{0.25cm}|}\hline -- & -- \tabularnewline \hline  + & + \tabularnewline \hline + & + \tabularnewline \hline  -- & -- \tabularnewline \hline  \end{tabular}
      \end{tabular}
      \hspace{.2in}
      \begin{tabular}{cc}
      \multicolumn{2}{c}{ \underline{Branch 4}}\\
\underline{Name} &  \underline{Sites}\\
\\
4o-horz &    \begin{tabular}{|>{\centering}p{0.25cm}|>{\centering}p{0.25cm}|}\hline -- & + \tabularnewline \hline  -- & + \tabularnewline \hline  \end{tabular}   \\
\\
6s &  \begin{tabular}{|>{\centering}p{0.25cm}|>{\centering}p{0.25cm}|}\cline{1-1} + & \multicolumn{1}{r}{ \ }  \tabularnewline \hline  -- & + \tabularnewline \hline -- & + \tabularnewline \hline  + &  \multicolumn{1}{r}{ \ }  \tabularnewline \cline{1-1}   \end{tabular}
      \end{tabular}
  \caption{Solutions $\phi^{(0)}_{n,m}$ to the stationary equation (\ref{stateq}) for $\mu = 1$ are listed with corresponding naming convention for each type.  Nonzero sites of the configuration are shown with the values $\phi^{(0)}_{n,m} = \pm 1$ denoted as plus or minus.  All other sites are zero.  Solutions are grouped by branch number, i.e. according to which merge after continuing in $\epsilon $.}
  \end{table}

  For $\epsilon > 0$ real-valued solutions $\phi^{(\epsilon )}_{n,m}$ are computed from $\phi^{(0)}_{n,m}\in\mathds{R}$ via  continuation in the coupling parameter $\epsilon$.  Such solutions satisfying the limit $\displaystyle\lim_{\epsilon\rightarrow 0} \phi^{(\epsilon)}_{n,m} = \phi^{(0)}_{n,m}$ are unique and guaranteed to exist for $\epsilon$ in some a neighborhood $I_0=(-\epsilon_0,\epsilon_0)$ by an application of the implicit function theorem.  From the stationary equation $F^{(\epsilon)}_{n,m}=0$ one can directly compute the solvability condition ${\rm Im}\left(\phi^{(\epsilon)}_{n,m}\overline{F}^{(\epsilon)}_{n,m}(\phi)\right) = 0$ where overline represents the complex conjugate.  That is, solutions $\phi^{(\epsilon)}_{n,m}$  are also roots of $\vec{g}=[g_{n,m}]$ for elements defined as $g_{n,m}=$
  \begin{equation}
 \epsilon {\rm Im}\left( \ \phi_{n,m}\left( \overline{\phi}_{n,m+1}+\overline{\phi}_{n,m-1}-\overline{\phi}_{n+1,m}-\overline{\phi}_{n-1,m}\right) \right).
\nonumber
  \label{gdef}
  \end{equation}
  Considered as a(n implicit) function of $\vec{\theta}$ and $\epsilon$, the vector function $\vec{g}$ can be expanded in Taylor series that is convergent on the interval $I_0$ \cite{dnls,ChowHale,GS}.  That is,
\begin{equation}
  \vec{g}(\vec{\theta},\epsilon) = \displaystyle\sum_{k=1}^\infty \epsilon^k \vec{g}^{(k)}(\vec{\theta}) \mbox{ where } \vec{g}^{(k)}(\vec{\theta}) = \frac{1}{k!}\partial^k_\epsilon \vec{g}(\vec{\theta},0).\label{taylor}
  \end{equation}

Since the initial configuration at $\epsilon = 0$ exhibits a gauge invariance $\vec{\theta} \rightarrow \vec{\theta} + \theta_0$ for $\theta_0\in\mathds{R}$ this gives a one parameter family of roots of $\vec{g}$ for any fixed $\epsilon\in I_0$.  This implies that if the first order Jacobian matrix $J=\partial \vec{g}^{(1)}/\partial \vec\theta$ has a simple zero eigenvalue, there exists a unique (modulo gauge transformation) analytic continuation of the limiting solution $\phi^{(0)}$ into the domain $I_0$ \cite{dnls,ChowHale,GS}.  Having provided the conditions for the existence of the different branches of solutions, we now turn to their corresponding spectral stability analysis.
  
\subsection{Spectral Stability}

For each example solution $\phi^{(\epsilon)}_{n,m}$ in Table I the stability is monitored for each fixed $\epsilon>0$ via the linearization ansatz
\begin{equation}
\label{pert}
u = e^{i\mu z}\left( \phi^{(\epsilon)}_{n,m} + \delta \left[ a_{n,m}e^{\nu z} + b^*_{n,m}e^{\nu^*z}\right] \right)
\end{equation}
which yields the order $\delta$ linear system
\begin{equation}
\left[ \begin{array}{cc} M_1 & M_2 \\ -M_2^* & -M_1^* \end{array} \right] \left[ \begin{array}{c} a\\ b \end{array} \right] = -i \nu \left[ \begin{array}{c} a\\ b \end{array} \right]\label{mat}
\end{equation}
where $M_1 = \epsilon \Delta  - \mu + 2|\phi^{(\epsilon)}|^2$ and $M_2 = \left(\phi^{(\epsilon)}\right)^2$.   Thus max(Re$(\nu)$) $> 0$ corresponds to instability,
yielding the relevant instability growth rate, while max(Re$(\nu)$) = 0 corresponds to (neutral) stability.  Note that $\psi = [a  \ b]^T$ represents a column vector of length $2N^2$, where $N\times N$ is the two-dimensional grid size.

In the numerical computations that follow, we identify the relevant solutions
via fixed point iterations and subsequently solve numerically the matrix
eigenvalue problem of Eq.~(\ref{mat}) to determine their stability.
However, it is particularly useful to have some theoretical prediction/expectation 
about which configurations should be expected
to be stable and which ones should not.  To that effect, we adapt
the methodology summarized in~\cite{dnls} (based on earlier works
such as~\cite{pkf1,pkf2}). This allows us to connect the stability
of the few-site configurations with the Jacobian of the solvability
conditions, as follows.

From equations (\ref{taylor}) and using the notation of Section A we may write the bifurcation function $\vec{g}^{(1)} $ as follows:
\begin{eqnarray}
\vec{g}^{(1)}(\vec{\theta}) &=& \vec{\delta}_{L}\sin(\vec{\theta} - \vec{\theta}_L)+ \vec{\delta}_{R}\sin(\vec{\theta} - \vec{\theta}_R)\\
&& - \vec{\delta}_{T}\sin(\vec{\theta} - \vec{\theta}_T) - \vec{\delta}_{B}\sin(\vec{\theta}_j - \vec{\theta}_B)\nonumber,
\end{eqnarray}
where we intend the equation to be considered element-wise in each of
the excited sites.

Thus the first order Jacobian matrix $J=\partial \vec{g}^{(1)}/\partial \vec\theta$ has entries  that can be computed manually given any example solution.  The diagonal vector of the matrix $J$ is $ \delta_{L}\cos({\theta} - {\theta}_{L})+\delta_{R}\cos({\theta} - {\theta}_{R})  - \delta_{T}\cos({\theta} - {\theta}_{T}) - \delta_{B}\cos({\theta} - {\theta}_{B})$.  Non-zero off-diagonal entries are of the form $\pm\cos(\theta_j-\theta_{\star,j})$ where the index of the nonzero entry is the index of nonzero entries of $\delta_\star$ for each of $\star = L,R,T,B$ with the plus sign corresponding to $T,B$ and the minus sign corresponding to $L,R$.  The eigenvalues $\lambda_i$ of $J$ are then connected to the full stability problem via the relation $\displaystyle\lim_{\epsilon\rightarrow 0} \nu_i^2/\epsilon = 2\lambda_i$, with the relevant proof going through in a
same way as with the elliptic case of~\cite{dnls}.

  \begin{table}
      \begin{tabular}{ccc}
\underline{Name} &  \underline{Jacobian $J$} & \underline{Eigenvalues $\{\lambda_i\}$}\\
\\
1s &   $\left( 0 \right)$ & $\{0\}$ \\
\\
2i-horz &  $ \left(\begin{array}{cc} 1 & -1 \\ -1 & 1 \end{array}\right) $&$ \{2,0\}$ \\
\\
 4o-vert & $ \left(\begin{array}{cccc} 2 & -1 & 0 & -1 \\ -1 & 2 & -1 & 0\\ 0 & -1 & 2 & -1\\ -1& 0 & -1 & 2\end{array}\right)$  &$\{4,2,2,0\}$ \\
\\
2o-horz & $ \left(\begin{array}{cc} 1 & -1 \\ -1 & 1 \end{array}\right) $&$ \{-2,0\}$  \\
\\
4o-line &   $ \left(\begin{array}{cccc} 1 & -1 & 0 & 0 \\ -1 & 0 & 1 & 0\\ 0 & 1 & 0 & -1\\ 0& 0 & -1 & 1\end{array}\right)$  &$\{2,\pm\sqrt{2},0\}$\\
\\
4i-sqr &   $ \left(\begin{array}{cccc} 0 & 1 & 0 & -1 \\ 1 & 0 & -1 & 0\\ 0 & -1 & 0 & 1\\ -1& 0 & 1 & 0\end{array}\right)$  &$\{\pm 2,0,0\}$ \\
\\
8s & $\left(\begin{array}{cccccccc} 2& -1& 0 & 0& 0& 0 &0 & -1\\ -1 & 1 & 1 & 0 & 0 & 0 & -1 & 0 \\ 0& 1 & 1 & -1 & 0 & -1 & 0 & 0\\
0& 0& -1 & 2 & -1 & 0 & 0 & 0 \\ 0& 0&0&-1 &2 & -1 & 0 &0 \\ 0 & 0& -1 & 0 & -1 & 1 & 1 & 0 \\ 0 & -1 & 0 & 0 & 0 & 1 & 1 & -1\\ -1 & 0& 0& 0& 0& 0& -1 & 2 \end{array}\right) $& $\begin{array}{c}\{4, 2 \pm \sqrt{2},  \\  \ \ \ \ \ \ 2, 2, \pm \sqrt{2}, 0 \}\end{array}$ \\
\\
4o-horz & $ \left(\begin{array}{cccc} -2 & 1 & 0 & 1 \\ 1 & -2 & 1 & 0\\ 0 & 1 & -2 & 1\\ 1& 0 & 1 & -2\end{array}\right)$  &$\{-4,-2,-2,0\}$ \\
\\
6s &  $\left(\begin{array}{cccccc} 1 & -1 & 0& 0& 0& 0 \\ -1 & -1 & 1 & 0 & 1 & 0 \\ 0 & 1 & -2 & 1 & 0 & 0\\ 0 & 0 & 1 & -2 & 1 & 0 \\ 0 & 1 & 0 & 1 & -1 & -1 \\ 0& 0& 0& 0& -1 & 1 \end{array}\right)$ &\hspace{-.2in}$\begin{array}{c} \{ -3.68133, -1.64207, \\  \ \ \ \ \  \  \pm\sqrt{3}, 1.3234, 0\} \end{array}$
      \end{tabular}
  \caption{The table lists the Jacobian matrix $J$ and corresponding eigenvalues $\{\lambda_i\}$ for each configuration. It is important to appreciate that each positive number for $\lambda_i$ on the right column translates into an unstable pair for the full problem eigenvalues $\nu_i$.}
  \end{table}

The Jacobian matrix for each example configuration that we consider is listed in Table II.  Based on the eigenvalues $\{\lambda_i\}$ listed in Table II the 1s, 2o-horz, and 4o-horz configurations are found to be stable for very small $\epsilon$.    For such small $\epsilon$ we find one unstable direction for configurations 2i-horz and 4i-sqr; two for 4o-line and 6s; three for 4o-vert; and six for 8s.  
{Note that adjacent in-phase excitations along the horizontal direction such as 2i-horz, 4i-sqr, and 8s lead to instability, as well as out-of-phase excitations in the vertical direction such as 4o-vert. } For more complex configurations
(like 6s or 8s),
whether or not they will bear an instability depends on whether they
include such unstable ``base ingredients'' i.e., any in phase pair
along the horizontal (as is the case for 8s) or out of phase pair
along the vertical (as is the case for 6s).
Having the analytical predictions of Table II at hand, we now turn to
a numerical exploration of the corresponding (potential) instabilities.


\section{Numerical Results: Existence, Stability and Dynamics}
  
\begin{figure}
\includegraphics[width=\columnwidth]{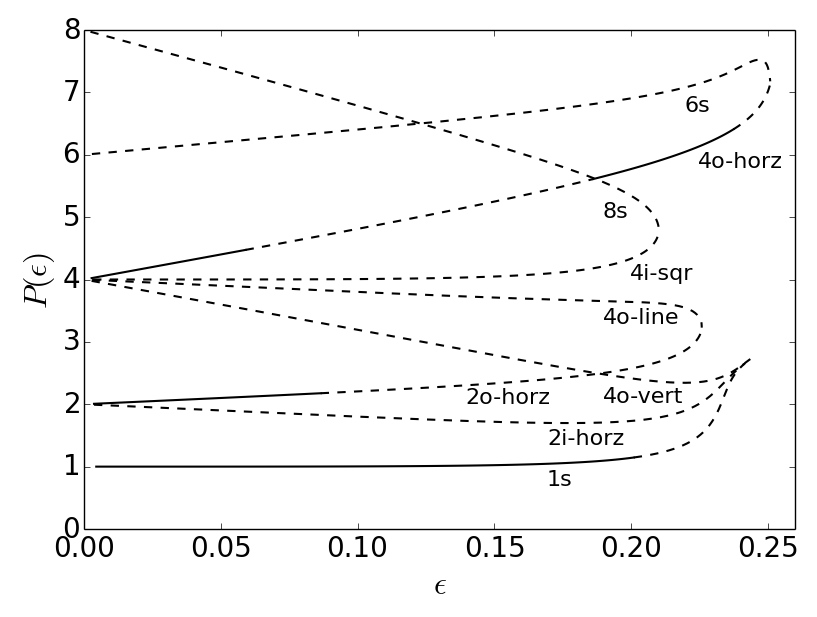}
\caption{The power  $P(\epsilon) = \sum_{n,m} |\phi^{(\epsilon)}_{n,m}|^2$ of solutions to the stationary equation (\ref{stateq}), obtained by arclength continuation, plotted as a function of $\epsilon>0$.  The continuation is initiated at $\epsilon = 0$ with the various solutions listed in Table I.  Underneath each branch segment is a label representing the $\epsilon=0$ configuration from which the solution is continued.  Dashed lines represent unstable solutions and solid
  ones represent stable solutions.  Branch 1 (lowest power) exists up to the value of $\epsilon \approx 0.242$ where solutions originating from types 1s, 2i-horz, and 4o-vert meet.  Branch 2 (lower middle) exists up to $\epsilon\approx 0.226$ at which point the solutions originating from types 2o-horz and 4o-line meet.  Branch 3 (upper middle) exists up to $\epsilon\approx 0.210$ where types 4i-sqr and 8s meet.  Branch 4 (highest power) exists up to $\epsilon \approx 0.251$ where types 4o-horz and 6s meet.
}
\label{pows}
\end{figure}

In the Appendix we provide a short pseudocode algorithm for the arclength continuation procedure that we utilize in order to identify the relevant branches
of solutions numerically.  The power of the resulting solutions $P(\epsilon) = \sum_{n,m} |\phi^{(\epsilon)}_{n,m}|^2$ is plotted as a function of $\epsilon > 0$ in Figure \ref{pows}.   The figure shows that the power curves merge into four
bifurcation ``events''.  The branch labels are indicated in the caption of Figure \ref{pows} as ordered from lowest to highest power.

\begin{figure}
\includegraphics[width=\columnwidth]{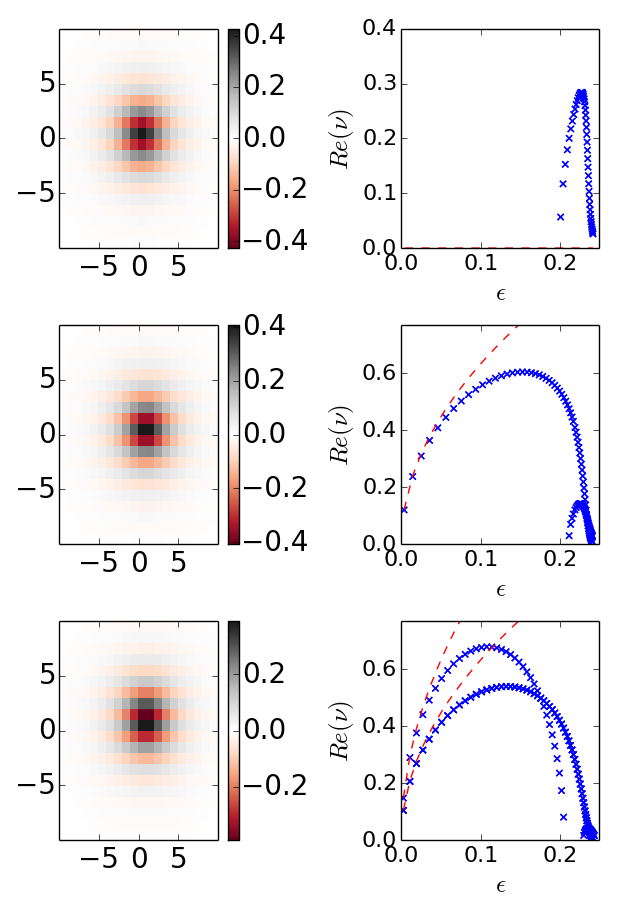} 
\caption{Each configuration plotted in the left column shows a Branch 1 (real-valued) solution that was obtained by continuation of the coupling parameter to the value of $\epsilon= 0.2414$ where the original configuration at $\epsilon = 0$ is 1s (top), 2i-horz (middle), or 4o-vert (bottom).  In the right column the set of values \{Re$(\nu) > 0$\} is plotted for the corresponding whole branch segment versus $\epsilon$ where similarly the original configuration at $\epsilon = 0$ is 1s (top), 2i-horz (middle), or 4o-vert (bottom).  For the right column, circles mark the values for which Im$(\nu)>0$ (i.e. $\nu$ is complex non-real and existing as quartets in the complex plane) and x's mark the values for which $\nu\in\mathds{R}$ (i.e. $\nu$ is real and existing as pairs on the real axis of the complex plane).  
  }
\label{newt1}
\end{figure}

\begin{figure}
\includegraphics[width=\columnwidth]{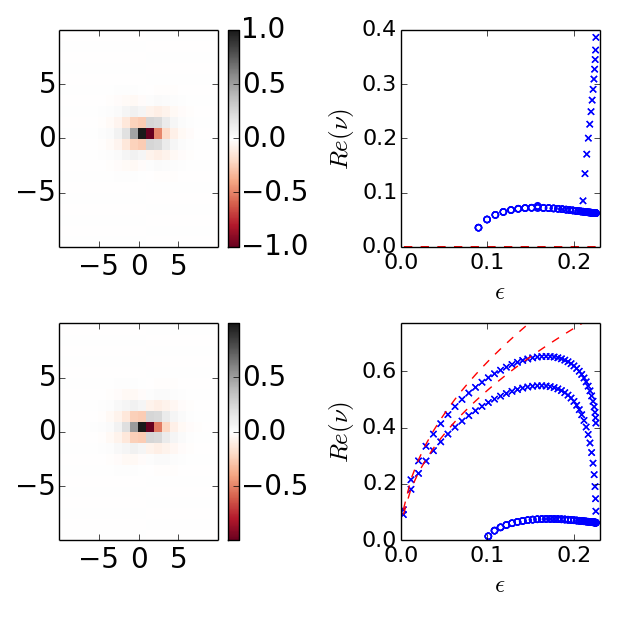}
\caption{Plots are similar to Figure \ref{newt1} but here for Branch 2 with the left column corresponding to $\epsilon=0.2258$ with initial $\epsilon=0$ configurations here as 2o-horz (top row) and 4o-line (bottom row).
}
\label{newt2}
\end{figure}

\begin{figure}
\includegraphics[width=\columnwidth]{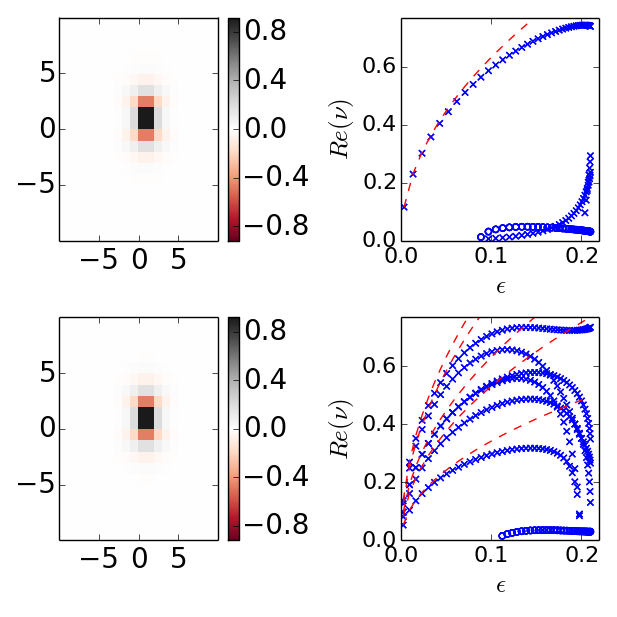} 
\caption{Plots are similar to Figure \ref{newt1} but here for Branch 3 with the left column corresponding to $\epsilon=0.2101$ with initial $\epsilon=0$ configurations here as 4i-sqr (top row) and 8s (bottom row).
 }
\label{newt3}
\end{figure}

\begin{figure}
\includegraphics[width=\columnwidth]{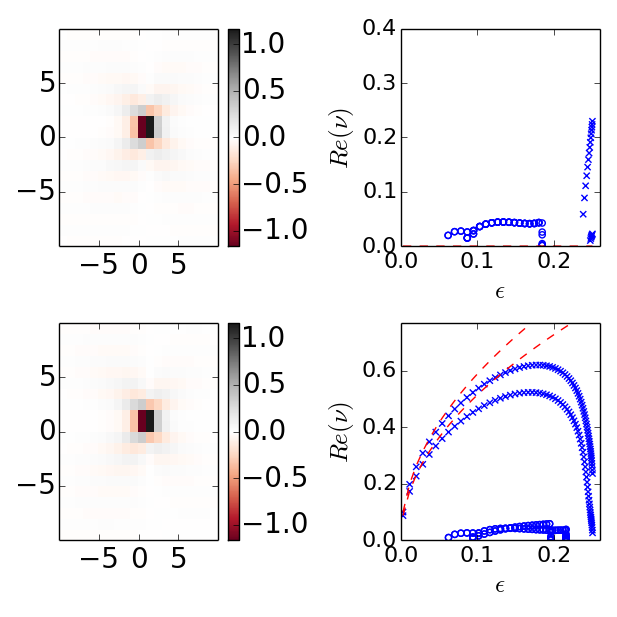} 
\caption{Plots are similar to Figure \ref{newt1} but here for Branch 4 with the left column corresponding to $\epsilon=0.2509$ with initial $\epsilon=0$ configurations here as 4o-horz (top row) and 6s (bottom row).
 }
\label{newt4}
\end{figure}

In Figures \ref{newt1}-\ref{newt4} the sample  solutions $\phi^{(\epsilon)}_{n,m}$ on the left columns show typical branch members
while the result of the two-dimensional continuation procedure over
$\epsilon$ is shown on the right panel through the unstable
eigendirection growth rates Re$(\nu)$. Comparing Table I to the left columns of Figures \ref{newt1}-\ref{newt4} we find that the extended solutions originate from the $\epsilon = 0$ solutions in the following manner as the nearest neighbor interaction is turned on for $\epsilon > 0$.  Generally sites to the left and right of the initial configuration become nonzero with the same sign as the initial configuration, and sites neighboring the initial configuration vertically become nonzero with opposite sign as the initial configuration.  Signs of the sites then alternate vertically and stay the same horizontally in a manner that respects the signs of the initial configuration as the footprint continues to expand for increasing $\epsilon > 0$.

Solutions initiated at $\epsilon = 0$ from the configurations types of 1s, 2i-horz, and 4o-vert merge into a single branch we denote as Branch 1.  These three solution types merge at $\epsilon\approx 0.242$ and the left column of Figure \ref{newt1} shows example solutions for $\epsilon = 0.2414$ on Branch 1.  In the right column of Figure \ref{newt1} we plot the nonzero real parts of eigenvalues $\nu$ as  computed from equation (\ref{mat}).  Note that the eigenvalue plots denote real eigenvalues with an ``x" mark and nonreal complex eigenvalues with an "o" mark.  For very small $\epsilon$ the prediction $\nu \approx \sqrt{2\lambda \epsilon}$ from Section B is plotted in a red dashed line based on the values of $\lambda$ in Table II. Clearly, the one unstable eigenvalue of
the 2i-horz configuration and the two unstable eigenvalues of
the 4o-vert configuration are well captured for small $\epsilon$.
As $\epsilon$ increases, however, these real eigenvalue pairs appear to
turn around towards $\nu=0$ and tend the origin of the spectral
plane as the bifurcation point is approached.

One can follow the eigenvalue diagrams in Figure \ref{newt1} with the following description of the change of the eigenvalue types as a function of $\epsilon$.  Starting with the 1s configuration, near the merging point there is one nonzero real pair of eigenvalues within the range $0.201\leq \epsilon \leq 0.241$.  { Near the merge point this one real pair of eigenvalues moves towards the origin.  Additionally two pairs of eigenvalues on the imaginary axis tend toward the origin (these are not reflected in Figure \ref{newt1}).}  The 2i-horz has one nonzero real pair for $0<\epsilon\leq 0.211$ and two real pairs for $0.212\leq \epsilon \leq 0.241$.  { Near the merge point the two real pairs and one additional imaginary pair approach the origin.}  The 4o-vert has initially a total of three nonzero pairs (one pair coinciding for a total of two distinct) within the range $0<\epsilon\leq 0.205$, then within the range $0.206 \leq \epsilon \leq 0.229$ two real pairs, and the remaining interval $0.230\leq \epsilon \leq 0.241$ again three nonzero pairs (one pair coinciding, two distinct).  { Nearest to the merge point the three real eigenvalues decrease in amplitude towards the origin.}

Solutions initiated at $\epsilon = 0$ from the configurations types 2o-horz and 4o-line merge at $\epsilon\approx 0.226$ as Branch 2.  Figure \ref{newt2} shows example solutions for $\epsilon = 0.2258$ in the left column.  Note that according to Table II the 2o-horz type is initially
(i.e., for small $\epsilon$) stable and the 4o-line type initially has two unstable directions for small $\epsilon$.  The predictions $\nu\approx\sqrt{2\lambda\epsilon}$ according to Table II are plotted as red dotted lines in the right column of Figure \ref{newt2}.
For small $\epsilon$, the two unstable eigendirections of 4o-line
are well captured. As $\epsilon$ is increased,
in the right column one can follow the eigenvalue changes over $\epsilon$.   The 2o-horz type has one quartet with nonzero real part within the range $0.083\leq \epsilon \leq 0.225$ and then within the range $0.210\leq \epsilon \leq 0.225$ there is an additional real pair.  The 4o-line type has two real pairs for $0<\epsilon \leq 0.225$ and an additional quartet for $0.101\leq \epsilon\leq 0.225$.  Near the merge point for Branch 2 the magnitude of the real parts of the quartet eigenvalues from both the 2o-horz and the 4o-line approach approximately $0.05$ as $\epsilon\rightarrow 0.226$.  The merge point also has one real pair with magnitude approximately $0.38$.  Notice that the 2o-horz type has the real pair {\it increasing} toward this value while the 4o-line type has its largest real pair {\it decreasing} toward this value (the smaller real pair for 4o-line type goes to zero). Moreover, to confirm the saddle-center nature of this bifurcation,
the 2nd real pair of the 4o-line branch decreases towards the origin $\nu=0$
as the bifurcation point is approached, while the 2o-horz branch has an
imaginary eigenvalue pair (not shown here)
tending to collide with this real pair (of 4o-line)
at the origin.

Solutions initiated from 4i-sqr and 8s merge at $\epsilon\approx 0.210$ as Branch 3 and Figure \ref{newt3} shows example solutions for $\epsilon = 0.2101$.  Note that according to Table II the 4i-sqr type has one unstable direction and 8s has six unstable directions for small $\epsilon$.  These predictions are plotted as red dotted lines in the right column of Figure \ref{newt3},
again in good agreement with the numerical results at least
for small values of $\epsilon$, before turning around towards $\nu=0$,
which in this case too happens around $\epsilon=0.1$.  The eigenvalue types change over $\epsilon$ as follows.  The type 4i-sqr has a real pair for $0<\epsilon\leq 0.210$, a second real pair for the interval $0.056\leq\epsilon\leq 0.210$ and a third real pair for the interval $0.203\leq\epsilon\leq 0.210$.  Note also that a quartet appears in the interval $0.088\leq\epsilon\leq 0.210$.  The type 8s has six real pairs for most of the $\epsilon$ range with two persisting for the smaller interval $0<\epsilon\leq 0.199$ and four persisting (two overlapping) for the whole interval $0<\epsilon\leq 0.210$. Additionally a quartet exists for $0.108\leq \epsilon\leq 0.210$.   On both types 4i-sqr and 8s the largest real pair approaches a magnitude of approximately $0.73$ as $\epsilon\rightarrow 0.2101$ and the complex quartet has the magnitude of its real part approaching 0.02.   On both types 4i-sqr and 8s, focusing on the smaller real pairs of eigenvalues, they approach a magnitude of either $0.24$ or $0.31$.  Notice that the 8s configuration has at this point two extra real pairs (the overlapping pair) unaccounted for thus far -- they approach zero as the merge point nears.
These are the eigenvalues responsible for this saddle-center bifurcation,
while the 4i-sqr branch has two corresponding pairs tending to $\nu=0$
from the imaginary side.

Solutions initiated from 4o-horz and 6s merge at $\epsilon\approx 0.251$ as Branch 4 in the final example among our saddle-center bifurcations.  Figure \ref{newt4} shows example solutions for $\epsilon = 0.2509$.  Table II predicts that 4o-horz is initially stable and 6s initially has two unstable directions.  The 4o-horz type has one complex quartet for $0.059\leq \epsilon \leq 0.082$ and three such for $0.083\leq \epsilon\leq 0.096$ (two coinciding), while two coinciding ones remain for $0.097\leq \epsilon\leq 0.185$.  {Near $\epsilon = 0.185$ these
  complex eigenvalues rapidly return to the imaginary axis.}  A real pair of eigenvalues exists
close to the merging point in the interval $0.237\leq\epsilon \leq 0.251$ and a second real pair appears very near that merge point for $0.248\leq\epsilon\leq 0.251$.  The 6s type has two real pairs of eigenvalues for $0 < \epsilon \leq 0.251$ and an additional quartet for $0.061\leq \epsilon\leq 0.091$ and a total of three quartets for $0.092\leq\epsilon\leq 0.102$; then, it has a total of two quartets for $0.103\leq \epsilon \leq 0.197$ and one quartet for $0.198\leq \epsilon\leq 0.217$.  {Near $\epsilon = 0.217$ the non-real quartets eigenvalues rapidly return to the imaginary axis.}  Near the merging point for Branch 4 the magnitudes of the two real pairs of eigenvalues {\it decrease} towards the limiting magnitude value of $0.02$ and $0.23$ as $\epsilon\rightarrow 0.251$ for the 6s configuration while the two real pairs of eigenvalues on the 4o-horz side of the branch {\it increase} towards those same magnitude values as $\epsilon\rightarrow 0.251$.
Thus, in the vicinity of this point, the two configurations collide
and merge through the associated turning point of this final
saddle-center bifurcation.

\begin{figure}
\includegraphics[width=\columnwidth]{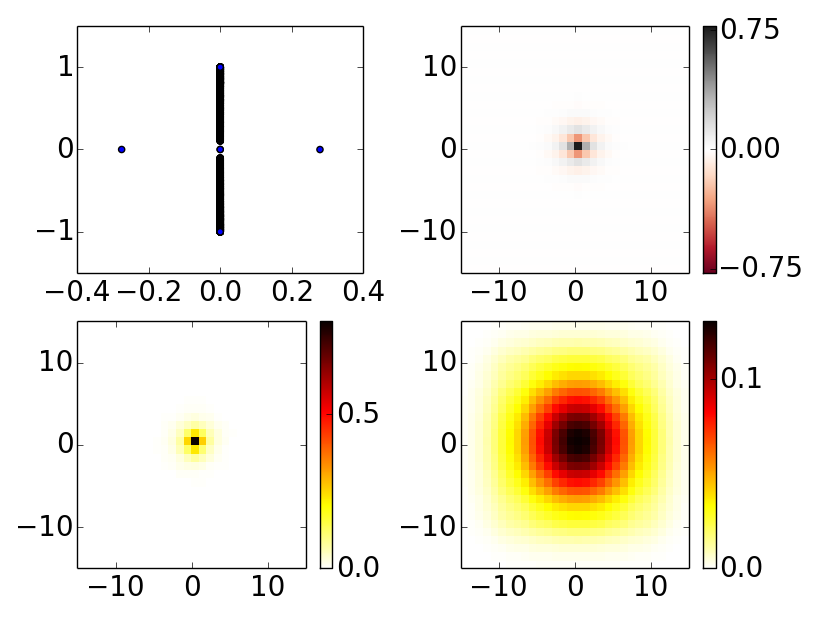}
\caption{The top left panel shows the values $\{\nu\}$, defined according to equations (\ref{pert}) and (\ref{mat}), plotted in the complex plane, for the corresponding stationary solution $\phi^{(\epsilon)}_{n,m}$ that is plotted in the top right panel.  The solution at $z=0$ is of type 1s for $\epsilon = 0.2250$ and it is real-valued.  The eigenvector $[a \ b]^T$ corresponding to the positive real $\nu$ value is used to perturb the solution according to equation (\ref{pert}) and $|a+b^*|$ is plotted in the bottom left panel.  After propagating according to equation (\ref{dyneq}) the result $|\phi^{(\epsilon)}_{n,m}(z)|$ is plotted in the bottom right for $z=50$.
}
\label{rk1}
\end{figure}

In Table III the dynamical fate of some case example solutions is listed together with the type of the perturbing eigenvectors. The 1s type solution at $\epsilon=0.2250$ evolves towards a single expanding mass marked as 1m in the table.  Figure \ref{rk1} shows the original solution at $z=0$ in the top right panel, the evolved solution at a later $z$ value in the bottom right panel, and the corresponding maximal eigenvalue and eigenvector in the top left and bottom left respectively.  The result of the evolution in the bottom right clearly illustrates the dispersive nature of the temporal dynamics.

\begin{figure}
\includegraphics[width=\columnwidth]{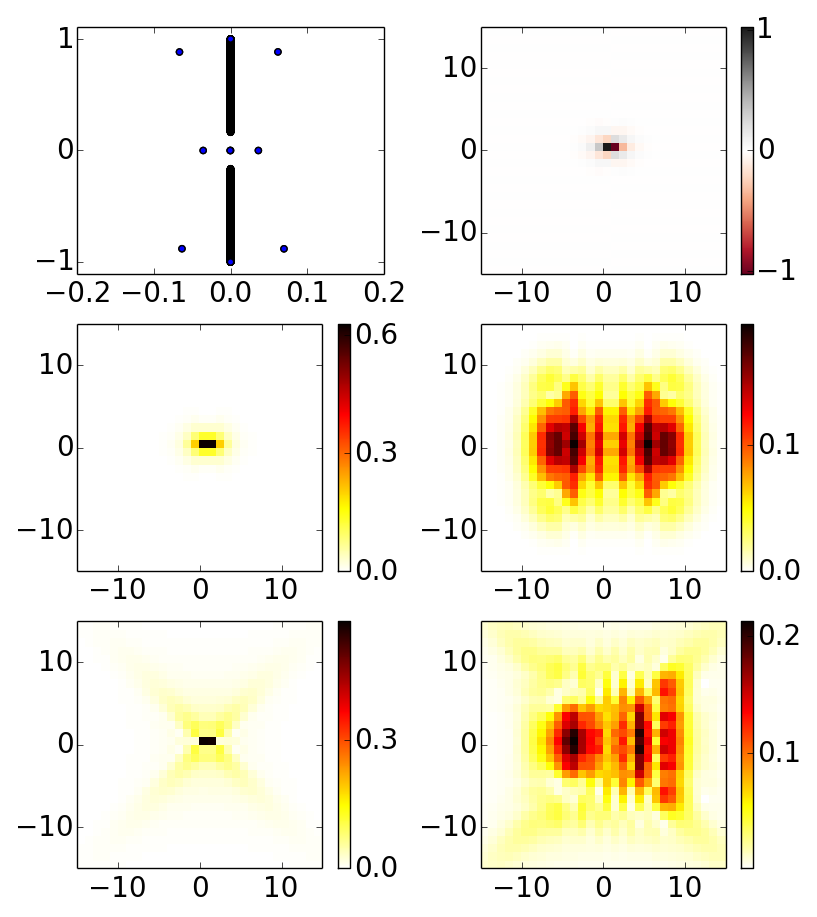}
\caption{Plots are similar to Figure \ref{rk1} where the top right panel shows the real-valued solution of type 2o-horz at $z=0$ for $\epsilon = 0.2096$.  The middle left panel shows $|a+b^*|$ for $[a \ b]^T$ the eigenvector corresponding to the eigenvalue with maximum real part among the quartet of eigenvalues (with Im$(\nu)\neq 0$).  Perturbing according to (\ref{pert}) with this eigenvector results in solution $|\phi^{(\epsilon)}_{n,m}(z)|$ in the middle right panel plotted at $z=218$.  The bottom row of panels is similar but for eigenvector corresponding to the real eigenvalue with maximum real part among the real eigenvalues (with Im$(\nu)=0$).  Perturbing according to (\ref{pert}) with this eigenvector results in solution $|\phi^{(\epsilon)}_{n,m}(z)|$ in the bottom right panel plotted at $z=122$. }
\label{rk2}
\end{figure}

Table III shows that at $\epsilon = 0.2096$ the solution of type 2o-horz has two different fates depending on whether one perturbs in the eigendirection corresponding to the maximal real eigenvalue (marked as the Im$(\nu)=0$ column of the table) versus the other eigendirection corresponding to the maximal complex eigenvalue (marked as the Im$(\nu)\neq 0$ column of the table).  In the former case the 2o-horz type solution evolves towards an expanding mass with two ``blobs" moving outwards along the horizontal direction; this is marked as 2m in the table and shown in the middle right panel of Figure \ref{rk2} with the corresponding eigenvector shown in the middle left panel.  In the latter case such expansion is not symmetric -- this is marked as 1-2m in the table and shown in the bottom right panel of Figure \ref{rk2} with the corresponding eigenvector shown in the bottom left panel.  Apparently here the perturbation added on top
of the initial 2o-horz configuration breaks its symmetry, leading to
the asymmetric evolution of the bottom right of Fig.~\ref{rk2}.

\begin{figure}
\includegraphics[width=\columnwidth]{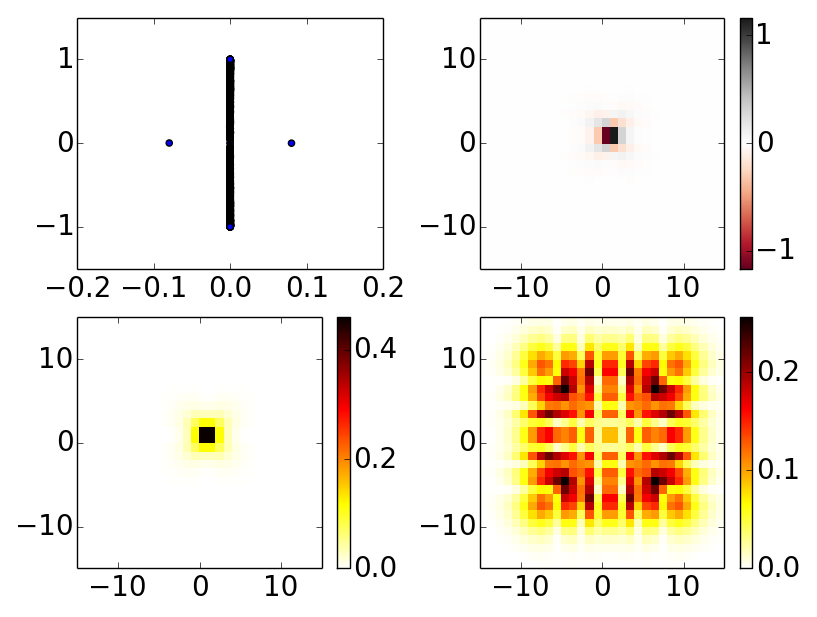}
\caption{Plots are similar to Figure \ref{rk1} where the top right panel shows the real-valued solution of type 4o-horz at $z=0$ for $\epsilon = 0.2400$.  The middle left panel shows $|a+b^*|$ for $[a \ b]^T$ the eigenvector corresponding to the eigenvalue with maximum real part.  Perturbing according to (\ref{pert}) with this eigenvector results in the solution $|\phi^{(\epsilon)}_{n,m}(z)|$ plotted in the bottom right panel at $z=116$.  }
\label{rk4}
\end{figure}

For other configurations such as 4o-line, 4i-sqr and 8s, according to 
Table III, their instabilities typically
led to a single-site resulting evolution
for the values of $\epsilon$-considered (for which the single site
configuration was dynamically stable).  {In the cases of 2i-horz and 4o-vert for $\epsilon=0.225$, the configurations approach a transient 1s state, vibrating near a 1s type solution with a pulsating core, since the stationary 1s
  configuration is unstable for this value of
  $\epsilon$.}
On the other hand, at $\epsilon = 0.2400$ the 4o-horz and 6s configurations evolve with mass expanding mostly towards the four corners, as is demonstrated in Figure \ref{rk4}.  Indeed, all other solutions shown in Table III
revert towards the 1s type, i.e., disperse mass while transforming
to a single site excited configuration.  We additionally tested solutions on Branches 1-4 when propagated according to equation (\ref{dyneq}) with $\epsilon$ beyond the bifurcation points such as $\epsilon = 0.28, 0.3$ to find that all tested standing wave solutions disperse for such higher $\epsilon$ values, closer to the continuum limit.  This is in line with the expectation that no
coherent structure exists in the vicinity of the continuum limit.
Yet, our quantitative analysis illustrates that dispersion dominates
already for rather weak couplings i.e., $\epsilon > 0.25$. Whether
a discrete analogue of self-similar dynamics arises for this
interval (corresponding to the continuum observations of~\cite{rumanov})
is an interesting open question for future study.

\begin{table}
\centering
\label{my-label}
\begin{tabular}{ >{\centering}p{1.5cm} | >{\centering}p{1.5cm} | >{\centering}p{1.5cm} | >{\centering}p{1.5cm} | >{\centering}p{1.5cm} | } \cline{2-5}
   & $Im(\nu) = 0 $  & $Im(\nu) \neq 0 $ & $Im(\nu) = 0 $ & $Im(\nu) \neq 0$  \tabularnewline\cline{2-5}
 & \multicolumn{2}{c|}{ $\epsilon = 0.1500$} &\multicolumn{2}{c|}{ $\epsilon = 0.2250$} \tabularnewline
  \hline
\multicolumn{1}{|c|}{ 1s } & (stable) &   --    & 1m &--  \tabularnewline 
\hline
\multicolumn{1}{|c|}{   2i-horz } &  1s  & --   & 1s-trans & --  \tabularnewline  
\hline
\multicolumn{1}{|c|}{  4o-vert } & 1s  &-- & 1s-trans & --  \tabularnewline 
 \hline
  \multicolumn{1}{|c|}{ }  & \multicolumn{2}{c|}{$\epsilon = 0.1500$} & \multicolumn{2}{c|}{$\epsilon = 0.2096$} \tabularnewline
  \hline
\multicolumn{1}{|c|}{ 2o-horz }  & -- & 1s  &2m  & 1-2m \tabularnewline 
\hline
\multicolumn{1}{|c|}{  4o-line } & 1s  &1s & 1s-trans &  1s-trans\tabularnewline 
\hline
  \multicolumn{1}{|c|}{ }  & \multicolumn{2}{c|}{$\epsilon = 0.1500$} & \multicolumn{2}{c|}{} \tabularnewline
  \hline
\multicolumn{1}{|c|}{ 4i-sqr }  & 1s   &1s  &  & \tabularnewline 
\hline
\multicolumn{1}{|c|}{ 8s } & 1s  & 1s & & \tabularnewline 
\hline
   \multicolumn{1}{|c|}{ } & \multicolumn{2}{c|}{$\epsilon = 0.1500$} & \multicolumn{2}{c|}{$\epsilon = 0.2400$ } \tabularnewline
 \hline
\multicolumn{1}{|c|}{  4o-horz }  &-- &1s & 4m & -- \tabularnewline 
\hline
\multicolumn{1}{|c|}{  6s }   & 1s & 1s & 4m & -- \tabularnewline  
\hline
\end{tabular}
\caption{The table lists the fate of solutions obtained via continuation in $\epsilon$ when they are propagated in the variable $z$ according to the dynamical equation (\ref{dyneq}).  Solutions are initiated at $z=0$ according to equation (\ref{pert}) with $\delta = 0.001$ and with eigenvector $[a \ b]^T$ corresponding to the eigenvalue $\nu$ that has largest real part where $\nu$ is either real (denoted as Im$(\nu)=0$) or in the complex plane with nonzero imaginary part (Im$(\nu)\neq 0$).  One circular mass is denoted as 1m such as the bottom right panel of Figure \ref{rk1}, two expanding masses is denoted as 2m such as the middle right panel of Figure \ref{rk2}, four masses expanding towards the corners of the grid is denoted as 4m such as the bottom right panel of Figure \ref{rk4}, and 1-2m corresponds to one mass expanding in one direction and a smaller mass in the opposite direction such as the bottom right panel of Figure \ref{rk2}.  {The abbreviation 1s-trans represents a transient state that is a pulsating 1s type.} }
\end{table}

\section{Conclusions \& Future Challenges}

In the present work, we have explored some of the fundamental solutions
of the hyperbolic discrete nonlinear Schr{\"o}dinger model. We have
initiated our search for such waveforms at the convenient anti-continuum
limit and have used continuation in the coupling parameter for some
of the most prototypical ones, most notably one-, two- and four-site
ones, with some exceptions of involving six- and eight-site ones,
when these were participating in bifurcations involving the lower
number of site branches. We have adapted the solvability condition
methodology of the elliptic case to this hyperbolic one and have
accordingly derived existence conditions and predictions for
the eigenvalues of the linearization of such few-site configurations.
Subsequently, we obtained the states via fixed point iterations
and examined the validity of the analytics as a function of the
coupling strength $\epsilon$. It was generally found that the
eigenvalue predictions worked well in the vicinity of the anti-continuum
limit. However, at larger values of the coupling (typically of
$\approx 0.2$), the eigenvalues were found in many configurations
to ``turn around'' and either meet up with a merging segment of the branch or return to the origin leading
to a set of bifurcation patterns that were elucidated herein,
some in fact involving
more than 2 configurations (as was the case with the branches 1s, 2i, 4o-vert).
This aligns itself with our expectation that {\it all} standing wave solutions
disappear
in the continuum limit~\cite{GHS3}. Sufficiently beyond these critical
bifurcation thresholds (all of which satisfied $0.2 \leq \epsilon_{cr}
\leq 0.25$ for the examples considered), the fate of standing wave-like
initial conditions was also examined and it was found that
they disperse, forming one or more dispersing ``blobs'', depending
on the form of the initial condition. Interestingly,
this type of fate (of dispersion into one or multiple blobs) could
arise for select examples before the bifurcation critical points,
as elaborated in Table III. Nevertheless, in numerous cases of
the latter scenario, the configurations just rearranged themselves
towards eventually reaching a single site state ($1s$).

Naturally, the present work paves the way for the numerous intriguing
questions both at the theoretical and at the numerical level. A difficult
set of questions concerns the phenomenology around $\epsilon_{cr}$.
Our analysis enables an understanding for small $\epsilon$; is there,
however, a way to capture the ``turning'' of the eigenvalues or
the emergence of these bifurcations around these critical values
of $\epsilon$~? Beyond these critical values, does one encounter
a discrete variant of the universal regimes presented in~\cite{rumanov}
and if so is there a way to analyze such phenomenology at the discrete
level~? Finally, extending considerations to the 3-dimensional setting
with two directions bearing the same sign of the dispersion (diffraction,
at the discrete level)
and one the opposite would be a possibility of interest in its own right.
Some of the questions are presently under consideration and will be
reported in future publications.

\vspace{5mm}

{\it Acknowledgements.} The authors acknowledge early efforts in this
direction by Dr. Kai Li. P.G.K. is also grateful to Profs. M.J. Ablowitz
and Boris A. Malomed for illuminating discussions on the subject.
This material is based upon work supported 
by the National Science Foundation under Grant No. DMS-1809074 (P.G.K.).

\section*{Appendix}
In Algorithm I an initial solution $\phi^{(0)}_{n,m}$ for $\epsilon_0 = 0$ from Table I is assumed to be represented as a column vector of length $N^2$, where $N\times N$ is the size of the two-dimensional grid.  The function $F(\phi, \epsilon) = F_\epsilon(\phi)$ is defined according to equation (\ref{stateq}) and also outputs a column vector of length $N^2$.  The constant values of the change in arclength parameter $ds$, the maximum $\epsilon$ value $\epsilon_{max}$, and the tolerance are assumed to be pre-set.
\begin{algorithm}[H]
\caption{Arclength Continuation}
\label{acont}
\begin{algorithmic}[1]
\State $\epsilon_0 = 0$; $\phi_0 = \phi^{(0)}$;
\State $v = $ nullspace$\left(\left[\frac{\partial F}{ \partial \phi}(\phi_0,\epsilon_0)  \ \ \frac{\partial F}{\partial \epsilon}(\phi_0);\right]\right)$;
  \State $v = v/$norm$(v)$;  \Comment{initial direction vector}
  \Do
\State $\phi = \phi_0; \epsilon = \epsilon_0;$
  \Do \Comment{Newton's method on augmented function $G$}
  \State $D = [\phi; \epsilon;] - [\phi_0; \epsilon_0;];$
  \State $G = [F(\phi,\epsilon);   \ D\cdot v-ds;]; $
  \State $M = \left[\frac{\partial F}{ \partial \phi}(\phi_0,\epsilon)  \ \ \frac{\partial F}{\partial \epsilon}(\phi); \ v^T;  \right]$
  \State $corr = M\backslash G$;
  \State $[\phi; \epsilon;] = [\phi; \epsilon;] -$ corr;
  \doWhile{norm(corr) $>$ tolerance}
\State $v = \left[\frac{\partial F}{ \partial \phi}(\phi_0,\epsilon)  \ \ \frac{\partial F}{\partial \epsilon}(\phi); \ v^T; \right] \backslash \left[ zeros(N^2,1); 1;\right] $;
  \State $v = v/$norm$(v)$;  \Comment{next direction vector}
\State $\phi_0 = \phi; \epsilon_0 = \epsilon;$
\doWhile{$\epsilon < \epsilon_{max}$}
\end{algorithmic}
\end{algorithm}

\end{document}